\documentstyle[preprint,aps]{revtex}
\begin{document}
\draft

\title{Kinetical foundations of non conventional statistics}
\author{G. Kaniadakis\footnote{electronic address: kaniadakis@polito.it}$^{,1}$,
P. Quarati\footnote{electronic addres: quarati@polito.it}$^{,2}$,
and A.M. Scarfone \footnote{electronic address:
scarfone@polito.it}$^{,3}$}
\address{\footnotesize $^{1,3}$Dipartimento di Fisica - Politecnico di Torino -
Corso Duca degli Abruzzi 24, 10129 Torino, Italy \\
Istituto Nazionale di Fisica della Materia - Unit\'a del
 Politecnico di Torino , and\\
 $^2$Istituto Nazionale di Fisica Nucleare - Sezione di
Cagliari, C.P. 170, I-09042 Monserrato, Italy}

\date{\today}
\maketitle

\begin {abstract}
After considering the kinetical interaction principle (KIP)
introduced in ref. Physica A {\bf296}, 405 (2001), we study in the
Boltzmann picture, the evolution equation and the H-theorem for
non extensive systems. The $q$-kinetics and the $\kappa$-kinetics
are studied in detail starting from the most general non linear
Boltzmann equation compatible with the KIP.\\

\end {abstract}

\section{Introduction}
It is common in equations of Boltzmann type to treat the
collisions as local and instantaneous. The spectral particle
function in energy and momentum is given by the Dirac function
$\delta(\epsilon-p^2/2\,m)$.\\ Assumptions to set Boltzmann
equation are based on three main features:
\begin{enumerate}
\item Triple collisions are neglected, BBGKY hierarchy is
truncated at the level of two-particle distribution function. One
is limited to study diluite gases of neutral particles with
short-range forces of interaction range much shorter than average
particle separation and mean-free particle path between
collisions. All the processes are governed by binary collisions;
\item Two-particle distribution function is factorizable with
one-particle distribution, i.e. two-body collisions are
statistically independent. One does not retain information from
previous encounters and memory about dynamical correlations;
Boltzmann equation can be solved with a class of dynamical
information.
\item The particle kinetics underlying the Boltzmann equation is
linear, i.e. the transition probability of a particle from one
site to another is given by factors proportional to the
distribution function $f$ rather than to powers of $f$ (as will be
discussed in the case of non linear kinetics).
\end{enumerate}
A non local and non instantaneous treatment of binary collisions,
as required in many different physical problems related, for
instance, to electron transport, Lorentz dense gas, rate
properties in non ideal plasma and heavy-ion reactions, is
strictly related to quasi-particle features and a Lorentz profile
$\delta_\gamma(\epsilon,\,p)$ of the spectral particle
distribution in energy and momentum can be assumed. A complete and
unique formulation and treatment of Boltzmann equation, in this
case, is not yet available, particularly for systems far from
equilibrium (which concept of quasi-particle works better in
kinetic equations for non equilibrium systems?).\\ With the
purpose of treating the Boltzmann equation in this context in the
near future, in the present work we limit ourselves to discuss a
generalization of the Boltzmann equation in order to treat
non-extensive systems composed by particles whose kinetics is non
linear and subject to binary, local and instantaneous
collisions.\\ This is achieved by employing the kinetic
interaction principle (KIP) which is local and instantaneous but
contains a generalized exclusion-inclusion principle which
eventually takes into account of a non linear kinetics. The KIP
has been introduced and described in refs. \cite{K1,K2}. It
permits to treat different statistical distributions already known
in literature in a unified way and in time dependent conditions.\\
The nonextensive statistics introduced by Tsallis \cite{Tsallis}
has been considered in time dependent conditions in the framework
of the Fokker-Planck pictures in refs. \cite{T1,T2,T6,T10,T11}.
Recently, the Tsallis kinetics has been studied also in the
Boltzmann picture in refs. \cite{K1,Lima}.\\ In the present
effort, starting from the KIP and after introducing the nonlinear
kinetics in the Boltzmann picture we consider in detail the
Tsallis statistics and the $\kappa$-statistics \cite{K1,K2,K3}

\section{Kinetical Interaction Principle}

Let us consider a particle system in the Boltzmann picture where
only point, binary collisions occur: $(A+A_1 \rightarrow
A^\prime+A^\prime_1)$. The most general non linear kinetics is
described through the evolution equation:
\begin{eqnarray}
&&\left[ \frac{\partial}{\partial t}+ \mbox{\boldmath
$v$}\frac{\partial }{\partial \mbox{\boldmath
$x$}}-\frac{1}{m}\,\frac{\partial V(\mbox{\boldmath
$x$})}{\partial \mbox{\boldmath $x$}}\,\frac{\partial }{\partial
\mbox{\boldmath $v$}} \right]\,f(t,\mbox{\boldmath
$x$},\mbox{\boldmath $v$}) \nonumber
\\ &&
= \int_{\cal R} d^nv^\prime\,d^nv_1\,d^nv^\prime_1\Big
[\pi(t,\,\mbox{\boldmath $x$},\,\mbox{\boldmath
$v$}^\prime\rightarrow\mbox{\boldmath $v$},\,\mbox{\boldmath
$v$}^\prime_1\rightarrow\mbox{\boldmath $v$}_1)
-\pi(t,\mbox{\boldmath $x$},\,\mbox{\boldmath $v$}\rightarrow
\mbox{\boldmath $v$}^\prime,\,\mbox{\boldmath $v$}_1\rightarrow
\mbox{\boldmath $v$}^\prime_1)\Big ] \ ,\label{1}
\end{eqnarray}
The transition probabilities are defined according to the {\it
Kinetical Interaction Principle} (KIP) \cite{K1,K2} by means of:
\begin{eqnarray}
\pi(t,\mbox{\boldmath $x$},\,\mbox{\boldmath $v$}\rightarrow
\mbox{\boldmath $v$}^\prime,\,\mbox{\boldmath $v$}_1\rightarrow
\mbox{\boldmath $v$}^\prime_1) = T(t,\,\mbox{\boldmath
$x$},\,\mbox{\boldmath $v$},\,\mbox{\boldmath
$v$}^\prime,\,\mbox{\boldmath $v$}_1,\,\mbox{\boldmath
$v$}^\prime_1)\,\gamma(f,\,f^\prime)\,\gamma(f_1,\,f^\prime_1) \
,\label{2}
\end{eqnarray}
being $f=f(t,\,\mbox{\boldmath $x$},\,\mbox{\boldmath $v$})$,
$f^\prime=f(t,\,\mbox{\boldmath $x$},\,\mbox{\boldmath
$v$}^\prime)$, $ f_1=f(t,\,\mbox{\boldmath $x$},\,\mbox{\boldmath
$v$}_1)$, $f^\prime_1=f(t,\,\mbox{\boldmath $x$},\,\mbox{\boldmath
$v$}^\prime_1)$ while the function $\gamma(f,\,f^\prime)$ assumes
the form:
\begin{equation}
\gamma(f,\,f^\prime)=a(f)\,b(f^\prime)\,c(f,\,f^\prime) \ .
\end{equation}

The first factor $a(f)$ is an arbitrary function of the particle
population of the starting site and satisfies the condition
$a(0)=0$. The second factor $b(f^\prime)$ is an arbitrary function
of the arrival site particle population obeying the condition
$b(0)=1$. The third factor $c(f,\,f^\prime)=c(f^\prime,f)$ takes
into account that the populations of the two sites, namely $f$ and
$f^\prime$, can eventually affect the transition, collectively and
symmetrically. \\The KIP imposes the following form to the
Boltzmann equation:
\begin{eqnarray}
&& \left[ \frac{\partial}{\partial t}+ \mbox{\boldmath
$v$}\frac{\partial}{\partial \mbox{\boldmath
$x$}}-\frac{1}{m}\,\frac{\partial V(\mbox{\boldmath
$x$})}{\partial \mbox{\boldmath $x$}}\,\frac{\partial}{\partial
\mbox{\boldmath $v$}}\right]\,f(t,\,\mbox{\boldmath
$x$},\,\mbox{\boldmath $v$})\nonumber \\ &&=\int_{\cal R}
d^nv^\prime\,d^nv_1\,d^nv^\prime_1\,T(t,\,\mbox{\boldmath
$x$},\,\mbox{\boldmath $v$},\,\mbox{\boldmath
$v$}^\prime,\,\mbox{\boldmath $v$}_1,\,\mbox{\boldmath
$v$}^\prime_1)\,c(f,\,f^\prime)\,c(f_1,\,f^\prime_1)\nonumber \\
&&\times\,\big [a(f^\prime)\,b(f)\,a(f^\prime_1)\,b(f_1)-
a(f)\,b(f^\prime)\,a(f_1)\,b(f^\prime_1)\big ] \ . \label{4}
\end{eqnarray}
After introducing the two functions:
\begin{equation}
B(f,\,f_1,\,f^\prime,\,f^\prime_1)=c(f,\,f^\prime)\,
c(f_1,\,f^\prime_1)\,b(f)\,b(f_1)\,b(f^\prime)\,b(f^\prime_1) \ ,
\end{equation}
and
\begin{equation}
\kappa(f)=\frac{a(f)}{b(f)} \ ,\label{6}
\end{equation}
one can write Eq. (\ref{4}) under the form:
\begin{eqnarray}
&&\left[\frac{\partial}{\partial t}+ \mbox{\boldmath
$v$}\frac{\partial}{\partial \mbox{\boldmath
$x$}}-\frac{1}{m}\,\frac{\partial V(\mbox{\boldmath
$x$})}{\partial \mbox{\boldmath $x$}}\,\frac{\partial }{\partial
\mbox{\boldmath $v$}} \right]\,f(t,\,\mbox{\boldmath
$x$},\,\mbox{\boldmath $v$})\nonumber \\&&=\int_{\cal R}
d^nv^\prime\,d^nv_1\,d^nv^\prime_1\,T(t,\,\mbox{\boldmath
$x$},\,\mbox{\boldmath $v$},\,\mbox{\boldmath
$v$}^\prime,\,\mbox{\boldmath $v$}_1,\,\mbox{\boldmath
$v$}^\prime_1)\,B(f,\,f_1,\,f^\prime,\,f^\prime_1)\nonumber
\\ && \times\,\Big \{\exp\,[\ln\kappa(f^\prime)+\ln\kappa(f^\prime_1)]-\exp\,[\ln
\kappa(f)+\ln\kappa(f_1)]\Big\} \ .\label{7}
\end{eqnarray}

We remark that this evolution equation describes a very large
class of non linear generalized kinetics which includes also the
standard linear kinetics obtained after posing $a(f)=f$,
$b(f^\prime)=1$ and $c(f,\,f^\prime)=1$.

We recall that in the frame of the present non linear kinetics the
mean value of a given quantity $A$ is defined through:
\begin{equation}
<A>=\int_{\cal R}d^nx\,d^nv\,A\,f \ \ \ ; \ \ \ \int_{\cal
R}d^nx\,d^nv\,f=1 \ .\label{Norma}
\end{equation}

\section{Generalized Entropy}

From Eq.(\ref{7}), in stationary conditions, we have:
\begin{equation}
\ln \kappa(f_{\!s})+ \ln \kappa(f_{\!1s}) =\ln
\kappa(f^\prime_{\!s})+\ln \kappa(f^\prime_{\!1s}) \ ,
\end{equation}
therefore the quantity $\ln \kappa(f_{\!s})$ is the collisional
invariant of the system. Taking into account that during the
collisions the particle number, energy and momentum are conserved,
we can  express the collisional invariant as \cite{K1}:
\begin{equation}
\ln\kappa(f_{\!s})=-\beta\,\left[\frac{1}{2}\,m\,v^2+
V(\mbox{\boldmath$x$})-\mu^\prime\right] \ ,
\end{equation}
with $\beta=1/k_{_B}\,T$. This last equation, which defines the
stationary distribution of the system, can be written also under
the form:
\begin{equation}
\kappa(f_{\!s})=\frac{1}{Z}\exp\left\{-\beta\,\left[\frac{1}{2}\,m\,\mbox{\boldmath$v$}^2+
V(\mbox{\boldmath$x$})-\mu\right]\right\} \ ,\label{11}
\end{equation}
with $\beta\,\mu=\beta\,\mu^\prime+\ln Z$ and $Z$ given trough
$\int d^nv\,d^nx\,f_{\!s}=1$.\\ Let us consider the functional
${\cal K}(t)$:
\begin{equation}
{\cal K}(t)= -k_{_B} \int_{\cal R}d^nx\,d^nv\int df\,\ln
\frac{\kappa(f)}{\kappa(f_{\!s})} \ . \label{12}
\end{equation}
Of course, the stationary distribution $f_{\!s}$ can be obtained
also by maximizing the functional ${\cal K}(t)$:
\begin{eqnarray}
&&\frac{\delta {\cal K}(t)}{\delta f}=0 \ \ \ \ \ \
\Longrightarrow \ \ \ \ \ \ f=f_{\!s} \ .\label{13}
\end{eqnarray}

In ref. \cite{K2} it has been shown that, if the condition:
\begin{equation}
\frac{d\kappa(f)}{df}\geq0
\end{equation}
is satisfied and $f$ obeys the Boltzmann equation (\ref{7}), the
quantity $-{\cal K}(t)$ is a Lyapunov functional:
\begin{eqnarray}
&&\frac{d{\cal K}(t)}{dt}\geq0 \ ,\label{15} \\ &&{\cal
K}(t)\leq{\cal K}(\infty) \ .\label{16}
\end{eqnarray}

In order to introduce the entropy $S(t)$ of the system, we pose:
\begin{equation}
{\cal K}(t)=S(t)-k_{_B}\,\beta\,(E-\mu^\prime) \ ,\label{17}
\end{equation}
and observe that the energy $E$ of the system:
\begin{equation}
E=\int_{\cal R}d^nx\,d^nv\,\left[\frac{1}{2}\,m\,
\mbox{\boldmath$v$}^2+V(\mbox{\boldmath$x$})\right]\,f\label{18}
\end{equation}
is a conserved quantity: $dE/dt=0$, as well as the particle number
$\int_{\cal R}d^nx\,d^nv\,f=1$. By comparing the two expressions
of the functional ${\cal K}(t)$ given by Eq.s (\ref{12}) and
(\ref{17}) we obtain that:
\begin{equation}
S= - k_{_B}\int_{\cal R}d^nx\,d^nv\,\int df \,\ln\kappa(f) \ .
\label{19}
\end{equation}
The entropy $S(t)$ obeys the H-theorem:
\begin{eqnarray}
&&\frac{dS(t)}{dt}\geq0 \ ,\label{20} \\ &&S(t)\leq S(\infty) \ ,
\end{eqnarray}
as can be verified immediately if we take into account Eq.s
(\ref{15})-(\ref{17}).

\section{Tsallis KINETICS}

Let us consider the  kinetics defined by fixing $\kappa(f)$
through:
\begin{equation}
\ln[Z\,\kappa(f)]= \ln_q(Z\,f) \ ,
\end{equation}
where $\ln_q(x)=(x^{1-q}-1)/(1-q)$ is the Tsallis logarithm and
$f$ is a normalized function: $\int_{\cal R}d^nx\,d^nv\,f=1$. The
evolution equation (\ref{7}) becomes:
\begin{eqnarray}
&&\left[\frac{\partial}{\partial t}+\mbox{\boldmath
$v$}\frac{\partial}{\partial\mbox{\boldmath
$x$}}-\frac{1}{m}\,\frac{\partial V(\mbox{\boldmath
$x$})}{\partial\mbox{\boldmath $x$}}\,\frac{\partial}{\partial
\mbox{\boldmath $v$}}\right]\,f(t,\mbox{\boldmath
$x$},\mbox{\boldmath $v$})=\int_{\cal R}\,d^nv^\prime\,
d^nv_1\,d^nv^\prime_1\nonumber
\\ && \times\,T(t,\,\mbox{\boldmath $x$},\,\mbox{\boldmath
$v$},\,\mbox{\boldmath $v$}^\prime,\,\mbox{\boldmath
$v$}_1,\,\mbox{\boldmath
$v$}^\prime_1)\,B(f,\,f_1,\,f^\prime,\,f^\prime_1) \nonumber \\ &&
\times
\left\{(Z^\prime\,Z_1^\prime)^{-1}\,\exp\,[\ln_q(Z^\prime\,f^\prime)
+\ln_q(Z^\prime_1\,f^\prime_1)]\right.\nonumber\\
&&\left.-(Z\,Z_1)^{-1}\,\exp\,[\ln_q(Z\,f)+\ln_q(Z_1\,f_1)]\right\}
\ ,\label{23}
\end{eqnarray}
where $Z,\,Z^\prime,\,Z_1$ and $Z_1^\prime$ are the partition
functions related to the distributions $f,\,f^\prime,\,f_1$ and
$f_1^\prime$ respectively.\\ Being $d\kappa(f)/df\geq0$ for the
system described by Eq. (\ref{23}), the H-theorem is satisfied and
its stationary distribution is given by:
\begin{equation}
 f_{\!s}=\frac{1}{Z}\exp_q\left\{-\beta\,\left[\frac{1}
 {2}\,m\,\mbox{\boldmath$v$}^2+
V(\mbox{\boldmath$x$})-\mu\right]\right\} \ ,\label{24}
\end{equation}
where $\exp_q(x)=[1+(1-q)\,x]^{1/(1-q)}$ is the Tsallis
exponential while the partition function is $Z=\int_{\cal R}d^nx\,
d^nv\,\exp_q\left\{-\beta\,\left[m\,
v^2/2+V(\mbox{\boldmath$x$})-\mu\right]\right\}$.

The above Eq. (\ref{24}) is the Tsallis distribution and can be
obtained also from the variational principle, defined by means of
Eqs. (\ref{12}) and (\ref{13}), which now assumes the form:
\begin{eqnarray}
\frac{\delta}{\delta f}\,k_{_B}\int_{\cal R}d^nx\,d^nv
\Bigg[-\frac{1}{2-q}\,\frac{(Z\,f)^{1-q}-1}{1-q}+\frac{1}{2-q}
-\beta \frac{1}{2}\,m\,v^2+\beta\,\mu \Bigg]\,f=0 \ .\label{25}
\end{eqnarray}
From Eq. (\ref{25}) it results clear that the entropy of the
system is given by:
\begin{equation}
S=-\frac{k_{_B}}{2-q}<\ln_q(Z\,f)>+\frac{k_{_B}}{2-q} \ ,
\end{equation}
and can be written also as:
\begin{equation}
S=\frac{1}{2-q}\,S^{^{\scriptscriptstyle(T)}}_{2-q}\,[Z\,f]+
\frac{k_{_B}}{2-q} \ ,
\end{equation}
being the Tsallis entropy defined through:
\begin{equation}
S^{^{\scriptscriptstyle(T)}}_{\, q}[f]=  -k_{_B}\int_{\cal
R}d^nx\,d^nv\,\frac{f-f^{\, q}}{1-q} \ .
\end{equation}

\section{THE $\mbox{\boldmath$\kappa$}$-KINETICS}

In ref. \cite{K1} it has been proposed the following one parameter
deformation of the logarithm function:
\begin{equation}
\ln_{_{\{{\scriptstyle \kappa}\}}}(x)=
\frac{x^{\kappa}-x^{-\kappa}}{2\,\kappa} \ ,
\end{equation}
which reduces to the standard logarithm as $\kappa\rightarrow 0$,
obeys the scaling law $\ln_{_{\{{\scriptstyle
\kappa}\}}}(x^m)=m\,\ln_{_{\{{\scriptstyle m\kappa}\}}}(x)$ and
presents the following  power law asymptotic behaviour:
\begin{eqnarray}
&&\ln_{_{\{{\scriptstyle
\kappa}\}}}(x){\atop\stackrel{\textstyle\sim}{\scriptstyle
x\rightarrow {\,0^+}}}-\frac{1}{|2\kappa|}\,x^{-|\kappa|} \ ,
\\ && \ln_{_{\{{\scriptstyle
\kappa}\}}}(x){\atop\stackrel{\textstyle\sim}{\scriptstyle
x\rightarrow +\infty}}\frac{1}{|2\kappa|}\,x^{|\kappa|} \ .
\end{eqnarray}
We consider the kinetics defined by fixing $\kappa (f)$ as
follows:
\begin{equation}
\ln[Z\,\kappa(f)]=\ln_{_{\{{\scriptstyle \kappa}\}}}(Z\,f) \ .
\end{equation}
The evolution equation (\ref{7}) becomes:
\begin{eqnarray}
&& \left[\frac{\partial}{\partial t}+\mbox{\boldmath
$v$}\frac{\partial}{\partial\mbox{\boldmath
$x$}}-\frac{1}{m}\,\frac{\partial V(\mbox{\boldmath
$x$})}{\partial \mbox{\boldmath $x$}}\,\frac{\partial}{\partial
\mbox{\boldmath $v$}}\right]\,f(t,\mbox{\boldmath
$x$},\mbox{\boldmath $v$})= \nonumber \\ && \int_{\cal R}
d^nv^\prime\,d^nv_1\,d^nv^\prime_1\,T(t,\,\mbox{\boldmath
$x$},\,\mbox{\boldmath $v$},\,\mbox{\boldmath
$v$}^\prime,\,\mbox{\boldmath $v$}_1,\,\mbox{\boldmath
$v$}^\prime_1)\,B(f,\,f_1,\,f^\prime,\,f^\prime_1) \nonumber
\\ &&\times\,\left\{(Z^\prime\,Z^\prime_1)^{-1}\,\exp\,[\ln_{_{\{{\scriptstyle
\kappa}\}}}(Z^\prime\,f^\prime)+\ln_{_{\{{\scriptstyle
\kappa}\}}}(Z^\prime_1\,f^\prime_1)]\right.\nonumber\\
&&\left.-(Z\,Z_1)^{-1}\,\exp\,[\ln_{_{\{{\scriptstyle
\kappa}\}}}(Z\,f)+\ln_{_{\{{\scriptstyle
\kappa}\}}}(Z_1\,f_1)]\right\} \ ,\label{33}
\end{eqnarray}
and reduces to the standard Boltzmann equation if we pose
$\kappa\rightarrow0$ and $B(f,\,f_1,\,f^\prime,\,f^\prime_1)=1$.
We observe that $d\kappa(f)/df \geq 0$ for $\forall \kappa \in
\mbox{\boldmath$R$}$ so that the H-theorem still holds.

The stationary distribution of Eq. (\ref{33}) is given by:
\begin{equation}
f_s = \frac{1}{Z}\exp_{_{\{{\scriptstyle \kappa}\}}}
\!\left[-\beta\,\left(\frac{1}{2}\,m\,v
^2+V(\mbox{\boldmath$x$})-\mu\right)\right] \ ,\label{34}
\end{equation}
with $Z=\int_{\cal R}d^nx\,d^nv\,\exp_{_{\{{\scriptstyle
\kappa}\}}}\left\{-\beta\,\left[m\,v^2/2+V(\mbox{\boldmath$x$})-\mu\right]\right\}$,
being the $\kappa$-exponential defined as the inverse function of
$\kappa$-logarithm:
\begin{equation}
\exp_{_{\{{\scriptstyle \kappa}\}}}(x)=
\left(\sqrt{1+\kappa^2\,x^2}+\kappa\,x\right)^{1/\kappa} \ .
\end{equation}
The $\kappa$-exponential reduces to the standard exponential as
$\kappa\rightarrow 0$, obeys the scale law
$[\exp_{_{\{{\scriptstyle
\kappa}\}}}(x)]^m=\exp_{_{\{{\scriptstyle \kappa/m}\}}}(m\,x)$ and
shows a power law asymptotic behaviour
\begin{equation}
\exp_{_{\{{\scriptstyle
\kappa}\}}}(x){\atop\stackrel{\textstyle\sim}{\scriptstyle
x\rightarrow \pm \infty}}|2\kappa\,x|^{\pm 1/|\kappa|} \ .
\end{equation}

It is easy to verify that the distribution (\ref{34}) can be
obtained after maximization under the appropriate constraint of
the entropy:
\begin{equation}
S_{\kappa}=-\frac{k_{_B}}{2\kappa}\,\int_{\cal R}d^nx\,d^nv
\,\left(\frac{Z^\kappa}{1+\kappa}\,f^{1+\kappa}-
\frac{Z^{-\kappa}}{1-\kappa}\,f^{1-\kappa} \right)\ ,
\end{equation}
which reduces to the standard Shannon entropy $S_0=-k_{_B}\int
d^nx\,d^nv\,f\ln(Zf)$ as the deformation parameter
$\kappa\rightarrow 0$. The entropy $S_{\kappa}$ and the entropy
$S$ defined in Eq. (\ref{19}) are connected through
$S_{\kappa}=S-k_{_B}\ln Z$. The variational equation (\ref{13})
reproducing the distribution (\ref{34}) can be written under the
form:
\begin{equation}
\frac{\delta}{\delta f}\,\bigg[S_{\kappa}-k_{_B}\,\beta\,E+
k_{_B}\,\beta\,\mu \bigg]=0 \ .
\end{equation}

\section{conclusion}
In the present contribution, we have studied the kinetics of a non
linear system in the frame of the Boltzmann picture. The main
points are the following:\\ The introduction of the KIP, given
through a particular expression of the transition probability
(\ref{2}), permits to describe the time evolution of a non linear
system and imposes, by means of the functional $\kappa(f)$ in Eq.
(\ref{6}), its steady state as stationary solution of the
evolution equation.\\ The KIP imposes the entropy form of the non
linear system. Its expression is given in Eq. (\ref{19}), and, as
shown in Eq. (\ref{20}), obeys to the H-theorem whenever  the
condition $d\kappa(f)/df\geq0$ is satisfied.\\ Finally, we have
discussed within the formalism here developed, as a working
example, the well known Tsallis statistics and the
$\kappa$-statistics, already introduced and described by one of us
in ref. \cite{K1}.

The evolution equation (\ref{23}) is very close to the equation
recently proposed by Lima, Silva, and Plastino \cite{Lima} within
a kinetic foundations of Tsallis thermostatistics. In some points
we differ from them. These authors consider a hard sphere particle
gas and a $q$-collisional term written in terms of a difference of
two $q$-exponent correlation functions (before and after
collision). Our collisional terms is a function deduced from KIP
and is a difference of two standard exponents.\\ In ref.
\cite{Lima} is reported that other possibilities also leading to
Tsallis distribution can be obtained if $\exp_q(x)$ is substituted
by other positive increasing functions $F_q(x)$ such that
$\lim_{q\rightarrow1}F_q(x)=\exp(x)$. An attempt to unify the two
derivations of Tsallis evolution equation is in progress.


\begin{thebibliography}{99}

\bibitem{K1}G. Kaniadakis, Physica A {\bf296}, 405 (2001).

\bibitem{K2}G. Kaniadakis, {\sl H-Theorem and Generalized Entropies
within the Framework of Non Linear Kinetics}, Phys. Lett. A (in
press); arXiv:cond-mat/0109192.

\bibitem{Tsallis} C. Tsallis, J. Stat. Phys. {\bf 52}, 279 (1988).

\bibitem{T1} A.R. Plastino, and A. Plastino, Physica A {\bf 222}, 347 (1995).

\bibitem{T2} C. Tsallis, and D.J. Buckman, Phys. Rev. E {\bf 54}, R2197 (1996).

\bibitem{T6} S. Martinez, A.R. Plastino, and A. Plastino, Physica A {\bf 259},
183 (1998).

\bibitem{T10} G. Kaniadakis, and G Lapenta, Phys. Rev. E {\bf 62},
3246 (2000).

\bibitem{T11} T.D. Frank, and Daffertshofer, Physica A {\bf 295}, 455 (2001).

\bibitem{Lima} J.A.S. Lima, R. Silva, and A.R. Plastino, Phys.
Rev. Lett. {\bf86}, 2938 (2001).

\bibitem{K3} G. Kaniadakis, and A.M. Scarfone, {\sl A New One
Parameter Deformation of the Exponential Function}, in the same
issue; arXiv:cond-mat/0109537.

\end{thebibliography}
\end{document}